\documentclass[sigconf]{acmart}
\AtBeginDocument{%
  }

\setcopyright{acmlicensed}
\copyrightyear{2024}
\acmYear{2024}
\acmDOI{XXXXXXX.XXXXXXX}

\usepackage{graphicx}
\usepackage{caption} 
\usepackage{subcaption} 
\usepackage{comment}

\begin{document}

\title{\textit{USM}: Unbiased Survey Modeling for Limiting Negative User Experiences in Recommendation Systems}



\author{Chenghui Yu}
\email{yuchenghui@tiktok.com}
\affiliation{%
  \institution{TikTok, Inc.}
  \country{USA}
  }

\author{Peiyi Li}
\email{lipeiyi.0@tiktok.com}
\affiliation{%
  \institution{TikTok, Inc.}
  \country{USA}
  }

\author{Haoze Wu}
\email{wuhaoze@tiktok.com}
\affiliation{%
 \institution{TikTok, Inc.}
 \country{USA}
 }  

\author{Yiri Wen}
\email{wenyr22@tsinghua.org.cn}
\affiliation{%
 \institution{TikTok, Inc.}
 \country{China}
 }   

\author{Bingfeng Deng}
\authornote{Authors corresponded for this research.}
\email{dengbingfeng@tiktok.com}
\affiliation{%
  \institution{TikTok, Inc.}
  \country{USA}
  }

\author{Hongyu Xiong}
\authornotemark[1]
\email{hongyu.xiong@tiktok.com}
\affiliation{%
  \institution{TikTok, Inc.}
  \country{USA}
  }

\renewcommand{\shortauthors}{Yu et al.}

\begin{abstract}
Reducing negative user experiences is essential for the success of recommendation platforms. 
Exposing users to inappropriate content could not only adversely affect users' psychological well-beings, but also potentially drive users away from the platform, sabotaging the platform's long-term success. 
However, recommendation algorithms tend to weigh more heavily on positive feedback signals due to the scarcity of negative ones, which may result in the neglect of valuable negative user feedback.
In this paper, we propose an approach aimed at limiting negative user experiences. 
Our method primarily relies on distributing in-feed surveys to the users, modeling the users' feedback collected from the survey, and integrating the model predictions into the recommendation system. We further enhance the baseline survey model by integrating the Learning Hidden Unit Contributions module and the Squeeze-and-Excitation module. 
In addition, we strive to resolve the problem of response Bias by applying a survey-submit model;
The A/B testing results indicate a reduction in survey sexual rate and survey inappropriate rate, ranging from -1.44\% to -3.9\%. 
Additionally, we compared our methods against an online baseline that does not incorporate our approach. The results indicate that our approach significantly reduces the report rate and dislike rate by 1\% to 2.27\% compared to the baseline, confirming the effectiveness of our methods in enhancing user experience.  After we launched the survey model based our approach on our platform, the model is able to bring reductions of 1.75\%, 2.57\%, 2.06\% on reports, dislikes, survey inappropriate rate, respectively. 
\end{abstract}




\maketitle

\section{Introduction}
Negative feedback signals are crucial to guardrail content recommendations and improve user experience. 
When these signals are effectively integrated into recommendation systems, they play a vital role in preventing the promotion of harmful or undesirable content, thereby contributing to a healthier online environment. 
However, the challenges associated with negative signals are noteworthy. Due to the limited visibility of options for users to express negative feedback, these signals are often sparse compared to positive signals. This imbalance can lead to a skewed understanding of user preferences, resulting in recommendations that prioritize short-term engagement over long-term satisfaction. 
Moreover, an over-reliance on positive signals can create a filter bubble, where users are continuously exposed to content that aligns with their immediate preferences but may not be beneficial in the long run. This scenario can ultimately lead to user attrition as audiences become disillusioned with the quality of the content provided.
Additionally, existing user signals frequently fail to meet specific customized requirements, such as understanding the underlying reasons for a user's likes or dislikes regarding a video. This lack of granularity hinders our ability to tailor content recommendations effectively, as we cannot identify the particular attributes of content that resonate with individual users. 

\begin{figure*}[h]
    \centering
    \begin{subfigure}[t]{0.23\textwidth}
        \centering
        \includegraphics[width=\linewidth]{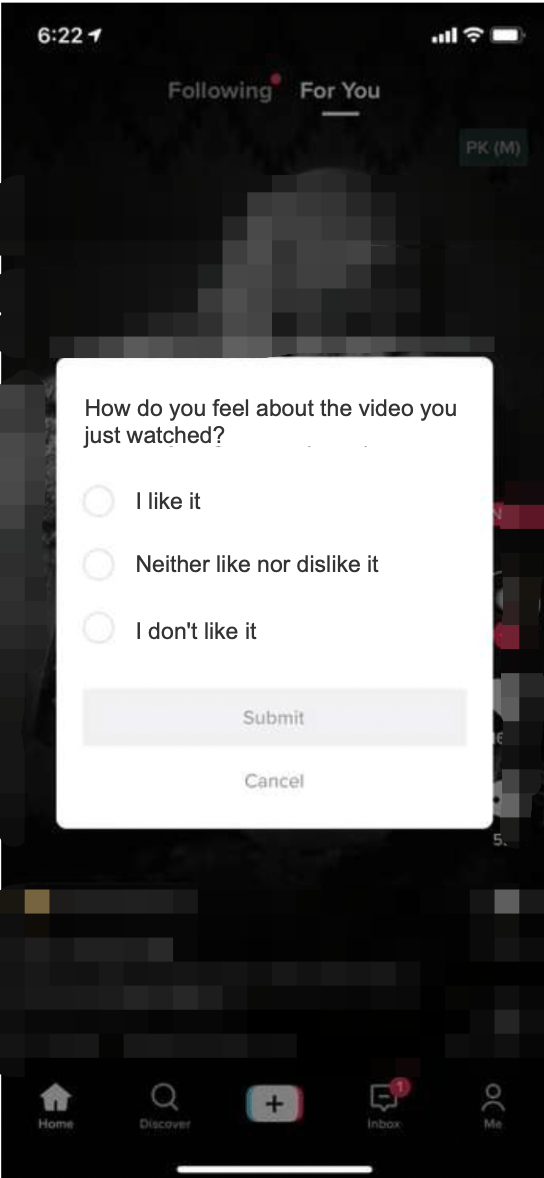}
        \caption{Satisfaction survey}
        \label{fig:img1}
    \end{subfigure}
    \hspace{1.5cm}
    \begin{subfigure}[t]{0.23\textwidth}
        \centering
        \includegraphics[width=\linewidth]{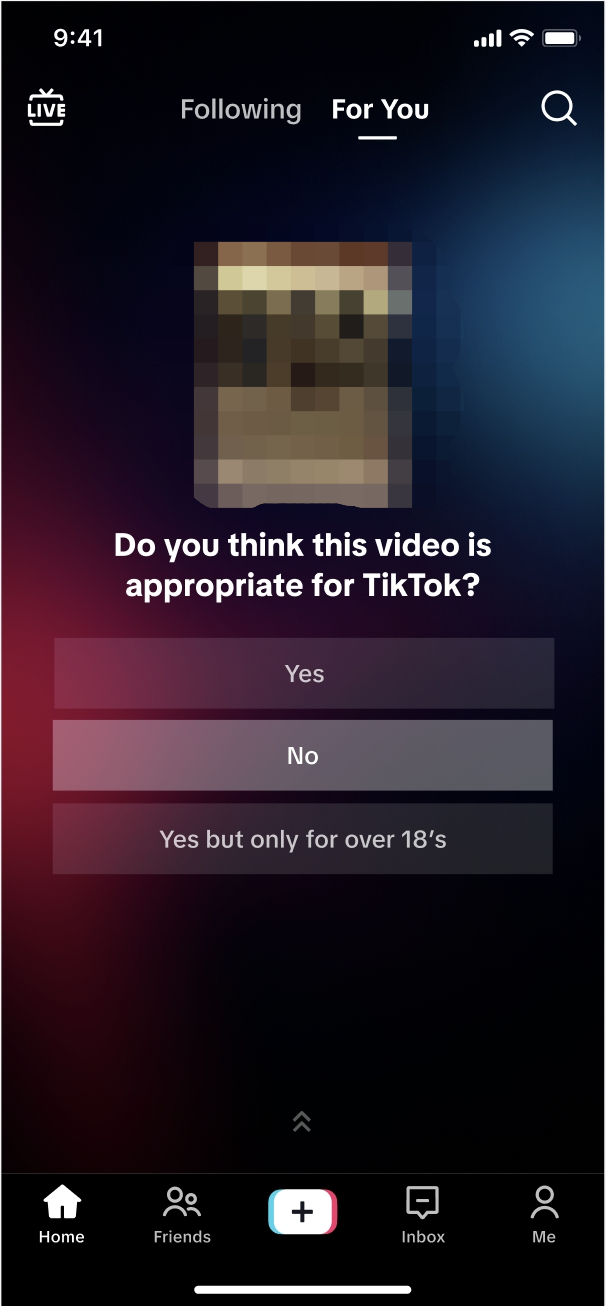}
        \caption{Inappropriate survey(primary page)}
        \label{fig:img2}
    \end{subfigure}
    \hspace{1.5cm}
    \begin{subfigure}[t]{0.23\textwidth}
        \centering
        \includegraphics[width=\linewidth]{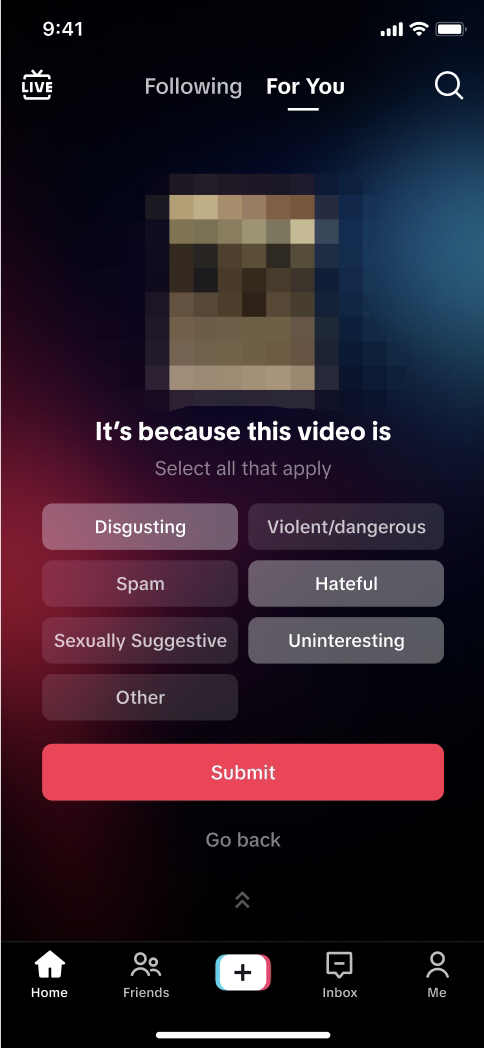}
        \caption{Inappropriate survey(secondary page)}
        \label{fig:img3}
    \end{subfigure}
    \caption{In-feed survey example. 
    (\subref{fig:img1}) illustrates the satisfaction survey designed to assess whether users appreciated the videos they recently viewed, offering three response options. 
    The subfigures (\subref{fig:img2}) and (\subref{fig:img3}) illustrate the primary and secondary pages of the inappropriate survey. 
    The primary page inquires whether the video is suitable on the TikTok platform; 
    If users select "no" or "only for 18+," a secondary interface is activated, prompting them to specify their reasons for the selection. 
        }
    \label{fig:three_images}
\end{figure*}

In this context, in-feed surveys can effectively address these limitations. 
By distributing surveys within users' for-you-feeds (FYFs), platforms can gather valuable insights into user opinions on specific issues. 
This approach enables platforms to gain a deeper understanding of user perceptions. We can utilize in-feed surveys for:

\begin{itemize}
    \item Monitoring content satisfaction and quality: This involves asking users about the quality and satisfaction of the recommended content, and using the survey results to help time-series monitoring and online A/B testing. 
    \item Optimizing recommendation algorithms: Building a deep learning model based on user responses for in-feed surveys, and utilizing the model prediction to further optimize users FYFs, thereby enhancing overall user experiences.
\end{itemize}

Currently, we primarily distribute two types of surveys on the platform, as illustrated in Fig \ref{fig:three_images}. 
One is the \textbf{Satisfaction Survey}, which inquires whether users liked the videos they recently viewed. 
We define the "survey like rate" to measure overall user satisfaction on the platform, as 
\begin{equation}
survey\_like\_rate = \dfrac{\sum_{m=1}^{M} like\_sub_{m}}{\sum_{n=1}^{N} all\_sub_{n} }
\label{eq.survey_like_rate}
\end{equation}
where $like\_sub$ represents a user submitting a survey with a "like" answer, and M denotes the total number of these users. $all\_sub$ collectively represents all individuals who submitted their results, while N denotes the total number of these individuals.
Our experiences have shown a strong correlation between the survey like rate and the number of daily active users (DAU).
Another survey we conduct is the \textbf{Inappropriate Survey}, which assesses users' perceptions regarding the suitability of the videos on our platform. It also asks for reasons behind any perceived inappropriateness. Potential reasons include being sexually suggestive, disgusting, hateful, violent, spammy, uninteresting, and other related concerns, each aligning with our trust \& safety (TnS) guidelines. Additionally, we define the "survey issue rate" to quantify the proportion of each issue identified in the survey, such as "survey sexual rate" and "survey inappropriate rate".
\begin{equation}
survey\_issue\_rate = \dfrac{\sum_{m=1}^{M} issue\_sub_{m}}{\sum_{n=1}^{N} all\_sub_{n} }
\label{eq.survey_issue_rate}
\end{equation}

The variable $issue\_sub$ refers to users who submit their results and report specific issues. Our experiences have shown a strong correlation between survey issues and the platform's TnS monitoring, which is enhanced by crowd-sourced annotations. The symbol M denotes the total number of users involved. The term $all\_sub$ represents all individuals who have submitted their results, while N signifies the total number of these individuals.

\begin{figure*}[h]
    \centering
    \includegraphics[width=0.85\textwidth]{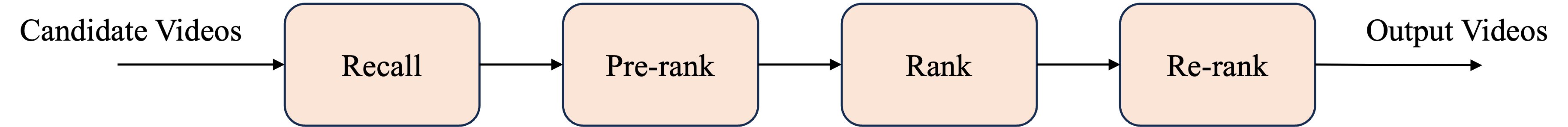}
    \caption{Illustration of the stages in a recommendation system. Recommendation systems can be roughly divided into four stages: recall, pre-rank, rank, and re-rank, ultimately selecting k videos (typically 8-10) as the final output for users.
    }
    \label{fig:re_system}
\end{figure*}

In this paper, we develop an in-feed survey model that utilizes user and item features as inputs, with survey responses serving as labels. 
We further enhance the baseline survey model by integrating the Learning Hidden Unit Contributions (LHUC) module and the Squeeze-and-Excitation (SE) module, which are utilized to selectively incorporate both universal and critical features, and to capture self-attention information, respectively.
We subsequently integrate this model into the recommendation system to optimize the users’ FYFs. 
This approach aims to enhance user satisfaction with the videos recommended by the platform, ultimately leading to improvement of daily active users (DAU) and corresponding TnS metrics. 
However, due to the sparse distribution of in-feed survey and users' different willingness to submit survey, we face significant challenges in terms of exposure bias and response bias during model training.
Correspondingly, we propose two approaches to address these challenges. 

The main contribution of this work can be briefly summarized as: 
\( 1) \) We distributed two types of surveys within FYF and utilized the survey responses to build the in-feed survey model, achieving significant gains in daily active users (DAU) and corresponding content issue metrics. 
\( 2) \) We developed a survey-submit model and used the estimated submission rates to weight the samples in the in-feed survey model, thereby resolving the issue of response bias.

\section{Related Work}

Our work is closely related to detection of negative user experiences, personalization models and solving problems of data bias in recommendation system.

\textbf{Negative Feedbacks.} The detection of negative user experiences in social media is a topic of significant interest, as evidenced by several studies analyzing these negative experiences \cite{gregoire2015managing, o2018social, o2020social}. 
To address the challenges of sparse and noisy negative feedback signals, certain models employ exposure variables \cite{liang2016modeling} or popularity metrics \cite{he2016fast} to effectively distill authentic negative signals from implicit feedback. 
Furthermore, some research \cite{xie2021deep, garcia2023detecting} focuses on developing personalized machine learning models aimed at predicting negative user experiences and applying these models within recommendation systems.

\textbf{Bias \& Debias.}In recent years, there has been significant growth in research on recommendation systems. Most studies focus on developing machine learning models to better analyze user behavior data. However, this data is typically observational rather than experimental. As a result, biases can easily be introduced, which we refer to as data bias.
The data collection process for recommendation systems is generally observational, not experimental. This means that sample selection and user decisions can be influenced by various undesirable factors, such as the exposure mechanisms of the recommendation systems or public opinions. Consequently, the distribution of training data can differ from the test data distribution. 
When the training data only captures a skewed snapshot of user preferences, the resulting recommendation model may yield suboptimal results. Therefore, data bias occurs when the distribution of the collected training data differs from the ideal test data distribution, exemplified by issues like Response Bias. \cite{marlin2012collaborative, hernandez2014probabilistic, steck2013evaluation} and Exposure Bias \cite{liu2020general, wang2016learning, chen2019samwalker, chen2020fast, zheng2021disentangling}.

\textbf{Response Bias.} Response Bias, also known as Selection Bias, occurs when users have the freedom to choose which items to rate. As a result, the observed ratings do not represent a true sample of all available ratings. In other words, the missing rating data is often categorized as Missing Not At Random (MNAR). A prior study conducted by Marlin et al. \cite{marlin2012collaborative} provides compelling evidence of the existence of selection bias in rating data. 
Specifically, they conducted a user survey to collect ratings for a set of randomly selected items and compared these to ratings for items chosen by users themselves. The presence of response bias is inherent in the data, leading to a distribution of observed ratings that differs from the overall distribution of all ratings \cite{hernandez2014probabilistic, steck2013evaluation}.
In the context of survey personalized modeling, response bias occurs when some users are more willing to participate in the survey than others. This discrepancy can result in training data that primarily consists of responses from users who are eager to engage. Currently, recommendation systems do not adequately address this issue, particularly in the realm of personalized modeling for surveys.

\section{Approaches}

In this section, we will first outline the structure of the baseline in-feed survey model and its application in recommendation systems. To tackle the challenges of exposure bias in survey modeling, we will introduce a survey-submit model.

\subsection{In-feed Survey Model}

Recommendation systems can be systematically divided into several stages: recall, pre-ranking, ranking, and re-ranking; as illustrated in Fig  \ref{fig:re_system}, later stages usually have a more direct impact on the final video recommendations. 
Currently, our survey model primarily focuses on the ranking stage, where it aims to reduce negative user experiences.

The model is trained using a dataset composed of user survey submission data, with different survey types distinguished by unique "survey-id" identifiers. 
For samples obtained from the satisfaction survey submissions, response of "I don't like it" are classified as positive examples, while all other responses are classified as negative examples. 
In the case of the inappropriate survey submissions, each head is associated with a specific option presented on the second page of the survey. Specifically, if a user selects an option, the sample is designated as a positive instance for the corresponding head; otherwise, it is designated as a negative instance.

\begin{figure}[ht]
    \centering
    \includegraphics[width=0.45\textwidth]{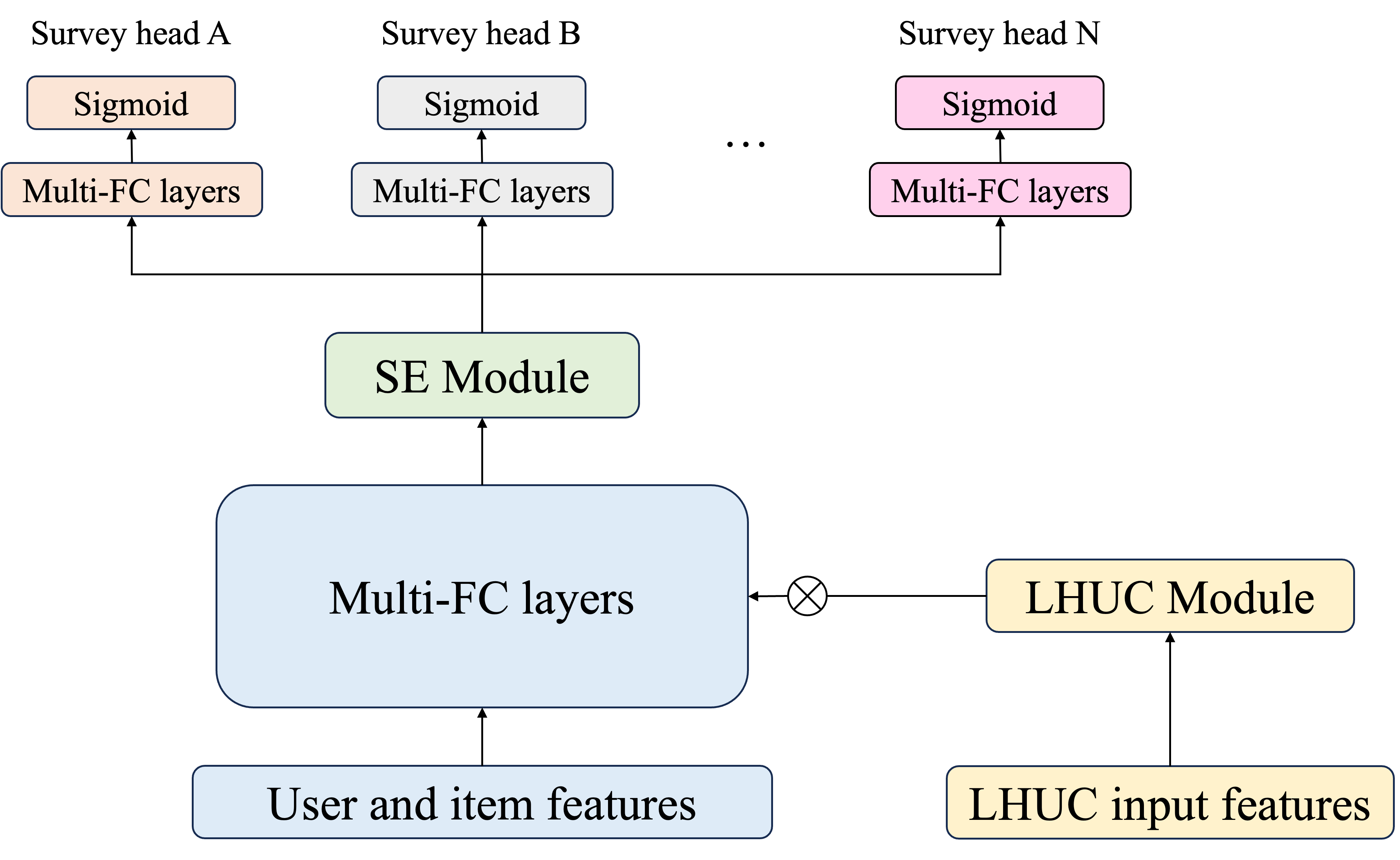}
    \caption{In-feed survey model structure. The model takes user and item features as inputs, along with a multi-head architecture where each head estimates the user's response for each specific option. We further enhanced the backbone by incorporating the LHUC and SE modules to improve feature cross-interaction and self-attention extraction.}
    \label{fig:in-feed_survey}
\end{figure}

The structure of the in-feed survey model is depicted in the Fig \ref{fig:in-feed_survey}. 
The architecture of the model is structured as a multi-head framework, in which each head estimates the probability of a user selecting a specific option. 
The backbone utilizes universal embeddings for users, items, and authors, along with additional raw attribute features of users and items as inputs, and consists of three fully connected (FC) layers with output dimensions of \{512, 256, 128\}.
Subsequently, we design a multi-head architecture, where each head comprises three FC layers with output dimensions of \{64, 16, 1\}. 

The multiple outputs from the In-feed survey model are combined with other scores (such as like and share) to calculate the final score for videos in the ranking stage. 
Ultimately, we pick the top-k (typically ranging from 8 to 10) highest-scoring videos to users. 
The item's final score formula can be expressed as follows:
\begin{equation}
final\_s(item) = \sum_{i=1}^{N} w_i \cdot p(survey\_i) + other\_s (item)
\label{eq.rank_score}
\end{equation}
where $final\_s$ denotes the final score of this item, $N$ denotes the number of survey model outputs, $w_i$ denotes the weight assigned to each survey head and is defined as a real-valued number, $p(survey\_i)$ denotes the predicted score of the $i$-th head in the survey model, and $other\_s$ denotes the scores of other heads, such as like score, follow score.

\subsection{LHUC Module \& SE Module}

We enhance the backbone of the model by incorporating the Learning Hidden Unit Contributions (LHUC) module and the Squeeze-and-Excitation (SE) module to facilitate feature cross-interaction and self-attention extraction, both of which have been substantiated as effective in numerous studies \cite{wu2019mutually, wu2021multi, zhu2020asta, wu2019cross}. The structures of the LHUC and SE modules are shown in Fig  \ref{fig:module}.

\subsubsection{LHUC Module}
The LHUC module \cite{swietojanski2016learning} aims at enhancing model performance by selectively incorporating both universal and critical features, improving feature interaction, and enabling the model to learn implicit information from both the original attributes of users and items and the inputs to the fully connected layers, ultimately improving personalized predictions and overall accuracy.

Universal features primarily encompass fundamental information about users, videos, and authors, such as user\_id, item\_id, and author\_id. Critical features, on the other hand, are determined through the calculation of feature importance, which quantifies the relevance of each feature in the context of the model's task. Feature importance is computed using a masking technique, where, during model evaluation, the corresponding feature is set to zero. The importance of a feature is inversely proportional to the decrease in model AUC: a larger reduction in AUC after masking a feature indicates a higher level of importance for that feature, as it signifies a more significant contribution to the model's predictive performance. The top-3 most important features selected for the LHUC module inputs are language, region, and device, which aligns with our intuition. All three features are closely related to the user’s context and significantly influence user preferences and behaviors. Language reflects the user’s communication preferences, region indicates geographic location and potential cultural influences, and device offers insights into the user’s platform of interaction. These contextual factors play a significant role in shaping user behavior, making them crucial for personalized predictions and recommendations.

The structure of the LHUC module is shown in  Fig  \ref{fig:LHUC}. The module employs input embeddings to generate three output embeddings through a series of fully connected layers, with dimensions of \{512, 256, 128\}, which are precisely aligned with the output dimensions of the multi-FC layers in the backbone network. The LHUC output embeddings \( LHUC_i \) are channel-wise multiplied with the outputs of the multi-FC layers \( FC\_\text{origin}_i \) to produce the final outputs \( FC\_\text{final}_i \), enabling effective feature modulation, as shown in Eq.\ref {eq.LHUC}:

\begin{equation}
FC\_\text{final}_i = LHUC_i \cdot FC\_\text{origin}_i  
\label{eq.LHUC}
\end{equation}
, where \( i \) corresponds to the number of layers in the multi-FC network.

\subsubsection{SE Module}
The SE module \cite{hu2018squeeze} is applied to the output of the final FC layer of the backbone, employing self-attention mechanisms to dynamically weight and recalibrate channel-wise features adaptively.

\begin{figure}[h]
    \centering
    \begin{subfigure}[t]{0.6\linewidth}
        \centering
        \includegraphics[width=\linewidth]{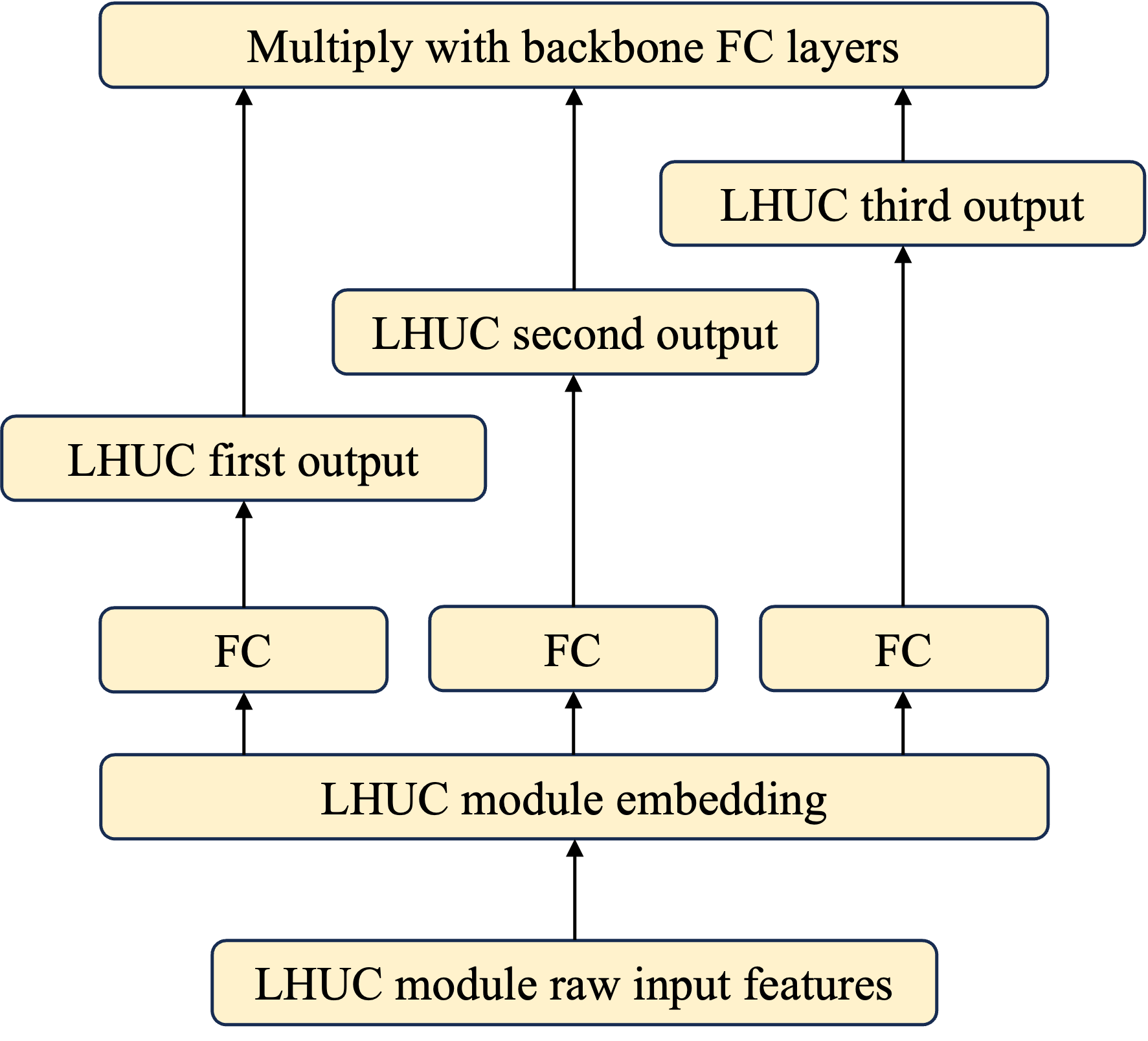}
        \caption{LHUC module}
        \label{fig:LHUC}
    \end{subfigure}
    \hfill
    \begin{subfigure}[t]{0.34\linewidth}
        \centering
        \includegraphics[width=\linewidth]{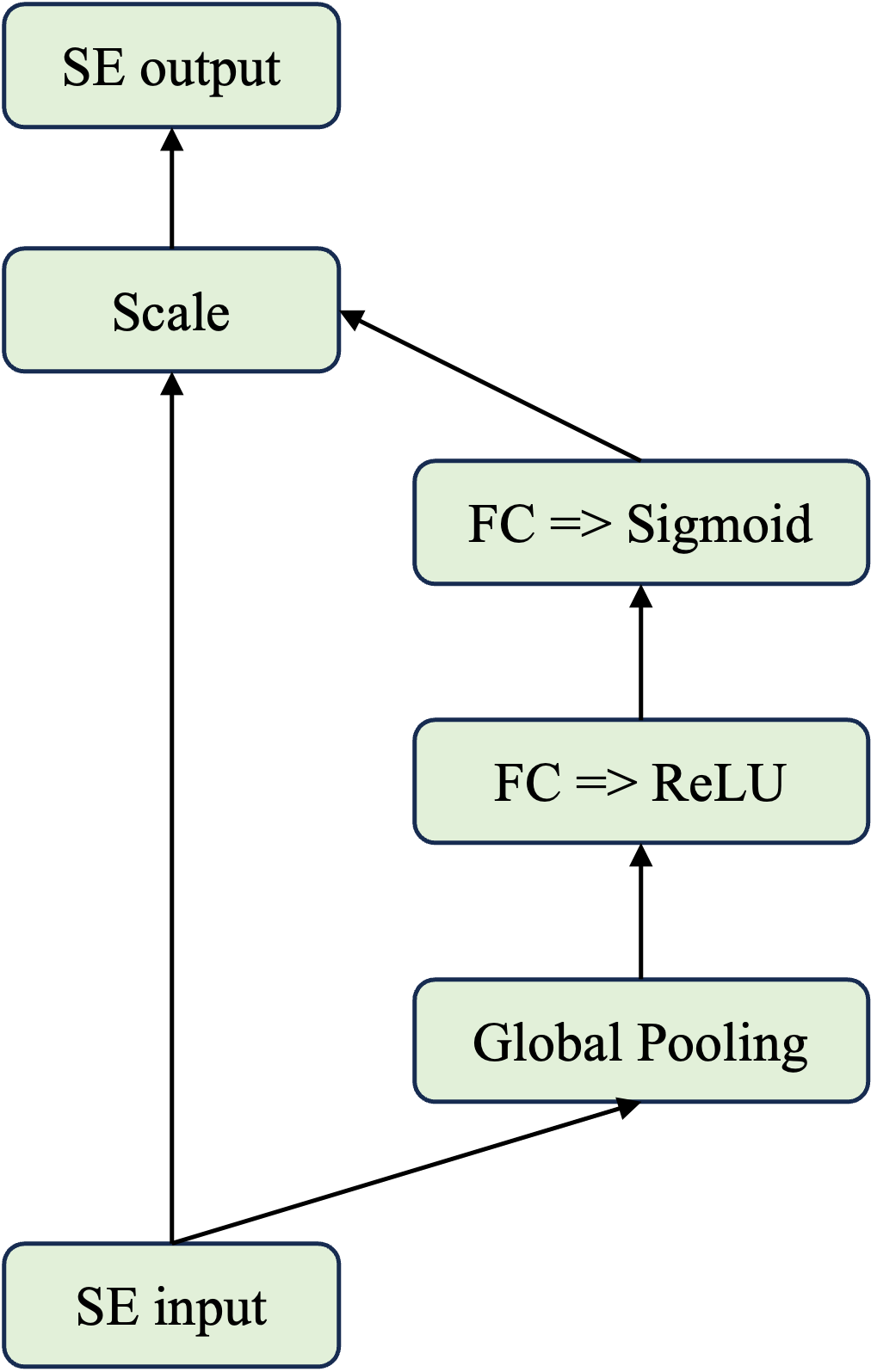}
        \caption{SE module}
        \label{fig:SE}
    \end{subfigure}
    \caption{ LHUC \& SE module detail structure. 
    (\subref{fig:LHUC}) The LHUC module utilizes the original input features to generate three output embeddings, which are multiplied with the outputs of the backbone FC layers;. 
    (\subref{fig:SE}) The SE module performs global pooling for squeezing, followed by two FC layers to extract self-attention, and multiplies it with the input for excitation.}
    \label{fig:module}
\end{figure}

The structure of the SE module is shown in  Fig  \ref{fig:SE}. The SE module initially applies a global pooling operation to squeeze the input along the channel dimension, producing a statistic \( \mathbf{s} \in \mathbb{R}^C \), where \( C \) represents the channel dimension. Subsequently, the SE module utilizes two fully connected layers to fully capture channel-wise dependencies \(e\), as expressed by Eq.\ref {eq.SE-1}: 

\begin{equation}
e = \sigma(W_2 \delta(W_1 \mathbf{s}))
\label{eq.SE-1}
\end{equation}
, where \(\delta\) refers to the ReLU active function \cite{nair2010rectified}, \(\sigma\) refers to the Sigmoid active function \cite{rumelhart1986backprop}, \(W_1 \in \mathbb{R}^{C/n \times C}\) and \(\quad W_2 \in \mathbb{R}^{C \times C/n}\). We set \( n=4\) to optimize the trade-off between computational efficiency and representation capacity, ensuring reduced resource consumption of the SE module while maintaining the integrity of inter-channel dependencies. The final output of the SE module is produced by rescaling the input using the activation \(e\):

\begin{equation}
output = F_{\text{scale}}(input, e) = e \cdot input
\end{equation}
, where \(F_{\text{scale}}\) refers to channel-wise multiplication. 

\subsection{Response Bias \& Survey-Submit Model}
After distributing the survey, we noticed varying levels of willingness among users to submit their responses. Some users are more active and interested in the questions, and after watching the video, they are more inclined to fill out the survey to provide feedback on their preferences. In contrast, other users are less enthusiastic and less willing to complete the survey they receive, which we called response bias. This indicates that their sensitivity to potential issues with the video differs. These two groups of users are distinct, and we need to address these differences in our client base to ensure the accuracy of our model.

\subsubsection{Problem Formulation}
We analyze response bias from a mathematical perspective, using the satisfaction survey as an example.
This survey primarily includes three questions: "I like it," "Neither like nor dislike it," and "I don't like it." When a user encounters a specific item, the probability formula for predicting whether they like that item is as follows:
\begin{equation}
P(like|ss) = P(like|ans)P(ans|ss)
\end{equation}
where $like$ represents the user selecting it in the survey, $ans$ represents the user answering and submitting survey results, $ss$ stands for survey show.

We could formulate an evaluation metric to assess the overall satisfaction levels on the platform, which is expressed as 
\begin{equation}
overall\_survey\_like\_rate = \dfrac{\sum_{i \in I} p(like|ss)}{|I|}
\end{equation}
, where $I$ represents all survey shows. If we assume that the user responding to the survey behaves in an unbiased manner, such that $p(ans|ss)$ is reduced to uniform distribution, we can use 
\begin{equation}
overall\_survey\_like\_rate = \dfrac{\sum_{i \in A} p(like|ans)}{|A|}
\end{equation}
, where $A$ represents all survey submits, and so it is the same formula as Eq.\ref{eq.survey_like_rate}.
We could always build a model to predict $\hat{p}(like|ans)$; however, users' willingness to answer and submit the survey is not uniform, thereby leading to response bias.

\subsubsection{Survey-Submit Model}

To address response bias during the training and evaluation phases of the prediction model $\hat{p}(like|ans)$ , we implement inverse propensity weighting (IPW) to ensure that each target item is weighted equally. 
This is accomplished by predicting a weight of $\dfrac{1}{P(ans|ss)}$ to every survey show
We call the method by which this is accomplished the survey-submit model.

\textbf{Model Architecture}
The structure of the survey-submit model is similar to in-feed survey model, though we don't include the LHUC module. 
Each head is associated with a specific type of survey and estimates the probability of a user submitting their answer to that survey. 
For instance, the "Satisfaction Survey Submit" head is used to predict the likelihood of a user submitting their response to the Satisfaction Survey after watching a video.

\textbf{Training Data}
We have collected various types of surveys online. When building the data flow, we consolidated all the survey submission information, including the satisfaction survey, sexual survey, and inappropriate survey. 
For model training, a positive sample is defined as a user who receives a survey and submits their answers.

\textbf{Features}
The features used in the model mainly include user features and item features.
User features include demographic information and generalized embeddings;  
Additionally, for the submit model, one of the most important features is the number of user history submissions.
Item features include generalized embeddings and counts and rates of any engagement actions applied to the item.

\section{Experiment}
\subsection{Metrics}
\subsubsection{AUC and Calibration}

For labeled samples, we utilize AUC (Area Under the Curve) to assess the offline model's ability to rank positive examples ahead of negative examples. 
Additionally, we use calibration values to measure the discrepancy between the model's average predictions and the actual event occurrence rates. The calculation method for calibration values is as follows:
\begin{equation}
Calibratoin =  avg(p_{u,i}) /avg(y_{u,i}) - 1
\end{equation}
Given that implementing A/B testing incurs costs, we will first compute AUC and calibration values offline to ensure that we can observe improvements in accuracy during model iterations. After launching the A/B testing, we will calculate online AUC and calibration values to ensure consistency between offline and online results.

\subsubsection{Neg/Pos-Feedback UAUC}
When the predicted models scores are used in the ranking phase, we place greater emphasis on the user-weighted average area under the curve (UAUC) metric. The UAUC is defined as an average of the area under the curve (AUC) for each user, aimed at assessing the model's ranking ability for different candidate items specific to individual users. However, in survey scenarios, the sparsity of feedback makes it challenging to collect responses from the same user with respect to multiple items, resulting in an inability to compute the UAUC metric for the survey behavior. However, we find a significant correlation between user reports (or dislikes) and responses indicating "inappropriate" content in surveys. Furthermore, since report (or dislike) buttons are exposed to all users, this leads to a denser collection of implicit feedback, allowing us to gather preferences for reports and dislikes from the same user across different candidate items. Therefore, we utilize reports and dislikes as implicit ground truths to evaluate the accuracy of the student model. 
We first calculate the Nfb (Negative-Feedback) AUC for each user, as 
\begin{equation}
\text{NfbAUC}_u = \frac{1}{P \cdot N} \sum_{i=1}^{P} \sum_{j=1}^{N} I(p_{i} > p_{j})
\label{eq.auc_per_u}
\end{equation}
where $I(p_i > p_j)$ is a metric function that takes the value of 1 when the prediction score of the positive sample $p_i$ is greater than that of the negative sample $p_j$; otherwise, it takes the value of 0. 
In this context, $P$ represents the number of videos exposed to the user and have been marked as disliked or reported, while $N$ represents the number of videos that have been exposed but neither marked as disliked nor reported.
Then, we average these values to obtain the Nfb UAUC, as shown in Eq.\ref{eq.user_uauc}, which better reflects the model's ability to identify negative feedback content across users:
\begin{equation}
\text{NfbUAUC} = \frac{1}{U} \sum_{u=1}^{U} \text{NfbAUC}_u
\label{eq.user_uauc}
\end{equation}
, where $U$ denotes number of users.

Similarly, for satisfaction survey, the Pfb (Positive-Feedback) UAUC metric is utilized to evaluate the performance of student model on surveys that have not been distributed. 
Specifically, in Eq.\ref{eq.auc_per_u}, $P$ is modified to represent the number of videos exposed to users that have been marked with positive interactions such as likes, shares, or favorites, while $N$ is modified to represent the number of videos that have been exposed to users but have not received any positive interactions.

\subsubsection{Online A/B Testing}

We launch the survey model on our online platform, where the model predictions will be applied to modify the ranking of candidates based on Eq.\ref{eq.rank_score}. 
The metrics for this A/B testing primarily focus on changes compared to the control group. 
We observe a reduction in negative issue feedback from users survey responses, and the metric for each issue is measured through Eq.\ref{eq.survey_issue_rate}.The issues may include inappropriate content, sexual content, dissatisfaction, etc. These metrics  are derived from actual user feedback obtained through survey results.

In addition, we also examine whether negative engagements (such as dislikes and reports) decline and whether positive engagements (such as likes and shares) increase. 
Additionally, we apply long-term holdout testing as a metric of improved user retention.

\subsubsection{Online Debiased A/B Testing}
In addition to the previously mentioned metrics, we have developed an online A/B testing metric to better assess the impact of model debiasing. This new metric is based on the survey-submit model for response bias. The specific calculation formula is provided below.
\begin{equation}
issue\_debias\_rate = \dfrac{\sum_{m=1}^{M} issue\_sub_{m}\times \dfrac{1}{sub\_pred_{m} } }{\sum_{n=1}^{N} all\_sub_{n}\times \dfrac{1}{sub\_pred_{n} }  }
\label{eq.survey_issue_rate_debias}
\end{equation}
 The variable $issue\_sub$ refers to users who submit their results and report specific issues, while $M$ denotes the total number of these individuals. $all\_sub$ collectively represents all individuals who submitted their results, while $N$ denotes the total number of these individuals. The variable $sub\_pred$ is the prediction of the survey-submit model.
\subsection{LHUC/SE Modules}
The experimental results show that integrating the LHUC/SE modules improves model performance across various surveys, as illustrated in Table \ref{tab:Neg/Pos-Feedback UAUC}. The AUC improvements indicate more accurate predictions of survey responses. Additionally, the negative/positive feedback UAUC metrics reveal that the model can better distinguish between individual users' positive and negative feedback on candidate items. This integration significantly enhances the model's ability to process both types of feedback, highlighting its superiority over the baseline model.

We also observed significant improvements in negative feedback metrics during our online A/B testing. Specifically, in the survey responses, we noted a 0.70\% decrease in the survey sexual rate and a 0.58\% reduction in the survey inappropriate rate. Regarding user-initiated negative feedback behaviors, the dislike rate decreased by 0.81\%. Through long-term reversal experiments, we found that these improvements led to a 0.01\% increase in long-term user retention, further validating the effectiveness and lasting impact of our approach.

\subsection{Survey-Submit Model}

We use the predicted score from the survey-submit model to debias the in-feed survey model training and evaluating which we called response debiased model. Table \ref{tab:Response Debias Model Online Metrics. AUC and calibration are calculated on labeled samples} shows the online AUC and calibration results after implementing response debiasing. 
For all three predicted targets, the treatment group demonstrates better performance compared to the control group. This suggests that this method allows us to more effectively model users who are reluctant to provide answers.

\begin{table}
    \centering
    \begin{tabular}{llll}
        \toprule
        Head& Model&  AUC&Calibration\\
         \midrule
         Satisfaction Survey& Baseline&  0.7416&0.06189\\
         & Response Debias&  0.7664&-0.0502\\
         \midrule
         Sexual Survey& Baseline&  0.7531&0.1465\\
         & Response Debias&  0.7808&0.02265\\
         \midrule
         Inappropriate Survey& Baseline& 0.7846&0.08077 \\
         & Response Debias&0.7963&-0.03104\\
         \bottomrule
    \end{tabular}
    \caption{Response Debias Model Online Metrics. AUC and calibration are calculated on labeled samples}
    \label{tab:Response Debias Model Online Metrics. AUC and calibration are calculated on labeled samples}
\end{table}

According to our findings, the survey-submit model indicates that users who are more willing to provide answers yield a higher estimated score, while those who are less willing result in a lower score. To further investigate this conclusion, we divided the predicted scores of the survey-submit model into different percentiles to compare the overall AUC and calibration performance of users who are unwilling to submit their survey answers against a baseline. Table \ref{tab:Separate Response Debias Model Online Metrics by Percentiles} presents the results of this analysis. It shows that, for users reluctant to submit their answers, the model's performance in the treatment group is superior to that of the control group. This outcome supports the effectiveness of our correction plan.

\begin{table*}
    \centering
    \begin{tabular}{llllc}
        \toprule
        Head & Percentiles & Model & AUC & Calibration\\
        \midrule 
         Satisfaction Survey & [0, P25)  &Baseline & 0.71  & -0.06883\\
                    &           &Response Debias& 0.855 & -0.0245\\
                   & [P25, P50)&Baseline & 0.6709  & 0.1613\\
                   &            &Response Debias& 0.7174 &0.008823\\
         \midrule
         Sexual Survey      & [0, P25)  &Baseline &0.75  & -0.743\\
                   &            &Response Debias&0.8667 &-0.4638\\
                   & [P25, P50)&Baseline &0.9182  &-0.6339\\
                   &            &Response Debias& 0.9412 &0.2484\\
         \midrule
         Inappropriate Survey&[0, P25)&Baseline &0.6719  &0.5715\\
                       &        &Response Debias&0.6973 &-0.2993\\
                      & [P25, P50)&Baseline & 0.6683  &-0.03581\\
                     &          &Response Debias&0.7245 &-0.01651\\
         \bottomrule
    \end{tabular}
    \caption{Separate Response Debias Model Online Metrics by Percentiles of Survey-submit Model Scores}
    \label{tab:Separate Response Debias Model Online Metrics by Percentiles}
\end{table*}

After we launched the model online, we assessed its real-world impact based on the actual user submission rates using online A/B test metrics. The results indicated that we achieved significant improvements across multiple online A/B tests, with a decrease of 1.44\% in the survey sexual rate and a 3.53\% reduction in the survey inappropriate rate. 

\begin{figure}
    \centering
    \includegraphics[width=1\linewidth]{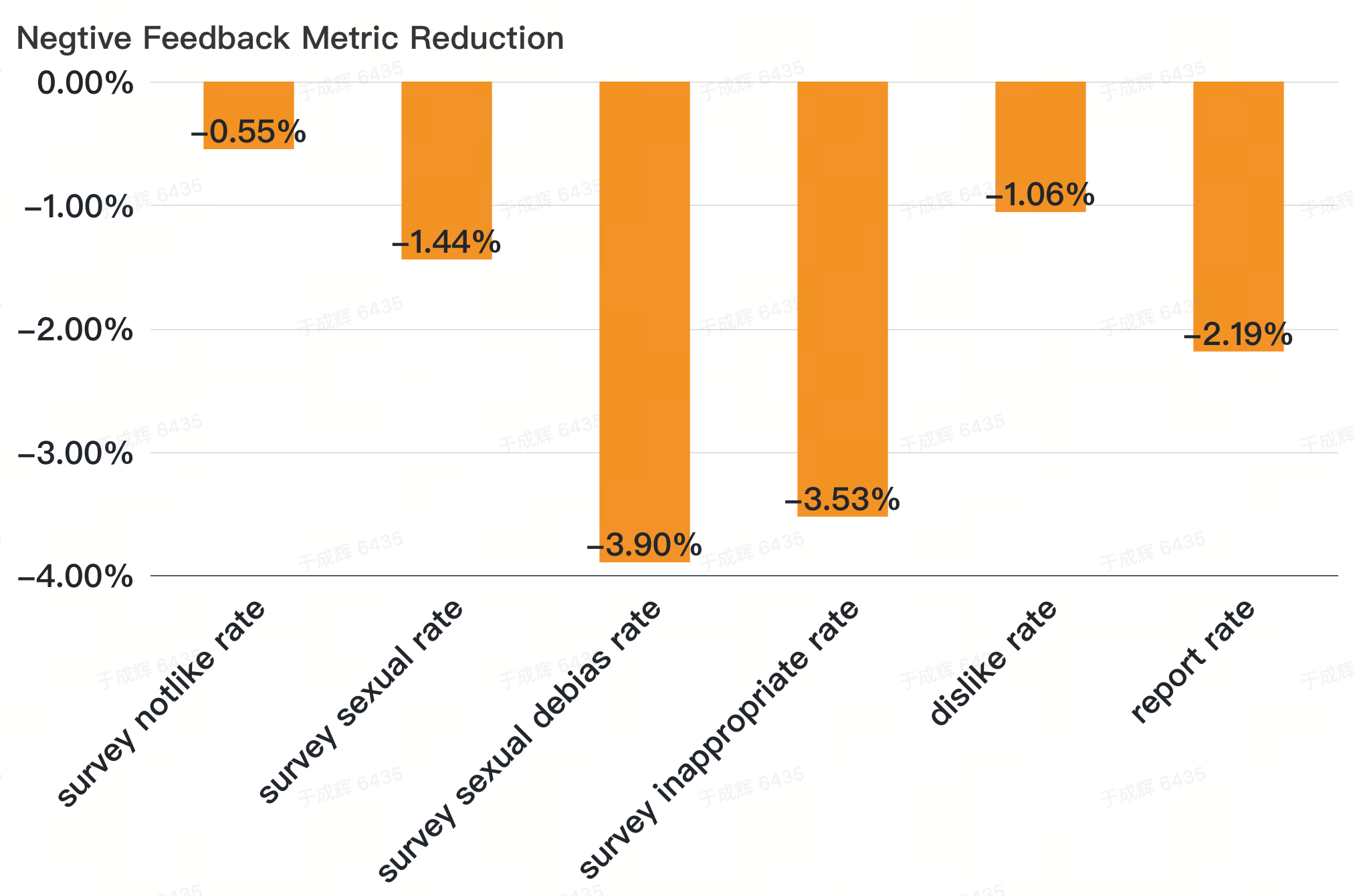}
    \caption{Negtive Feedback Metric Reduction using Response Debias}
    \label{fig:Negtive Feedback Metric Reduction using Response Debias}
\end{figure}

Importantly, we evaluated the effect of the response debiasing model on the adjusted online A/B testing metrics which are based on Eq. \ref{eq.survey_issue_rate_debias}. As illustrated in Fig \ref{fig:Negtive Feedback Metric Reduction using Response Debias}, the corrected metric for the survey sexual debiased rate showed that the model reduced this rate by 3.9\% compared to the uncorrected metrics. This reduction exceeds the previous decrease, validating the effectiveness of our method.

To further validate the effectiveness of our approaches, we combined LHUC and SE block  with the survey-submit model. Specifically, we used the predicted scores generated by the survey-submit model to reduce response bias in survey model, addressing response bias. After deploying the trained model on the online platform, we observed that its online UAUC was higher than that of the model that relied solely on LHUC and SE block, as indicated by online predictive accuracy metrics. For more details, please refer to Table \ref{tab:Neg/Pos-Feedback UAUC}, which illustrates the effectiveness of our approach.

\begin{table}
    \centering
    \begin{tabular}{llll}
        \toprule
        Head& Model&   AUC&UAUC\\
         \midrule
         Satisfaction Survey& Baseline&   0.7416&0.7000\\
         & LHUC&   0.7720&0.7123\\
         & LHUC+Debias& 0.7722  &0.7173\\
         \midrule
         Sexual Survey& Baseline&   0.7531&0.6678\\
         & LHUC&   0.7956&0.6759\\
         & LHUC+Debias& 0.7966  &0.6810\\
         \midrule
         Inappropriate Survey& Baseline&  0.7846&0.6990\\
         & LHUC& 0.8040&0.7003\\
         & LHUC+Debias&  0.8061 &0.7051\\
         \bottomrule
    \end{tabular}
    \caption{AUC and Neg/Pos-Feedback UAUC. Satisfaction Survey is evaluated by Pos-Feedback UAUC, Sexual Survey and Inappropriate Survey is evaluated by Neg-Feedback UAUC}
    \label{tab:Neg/Pos-Feedback UAUC}
\end{table}

\section{Conclusion}
We studied the problem of providing a better content recommendation on TikTok platform, focusing on limiting negative user experiences. 
The approach we proposed consists of a personalized, response-debiased, and exposure-debiased survey modeling framework. 
In addition, we proposed an evaluation framework consisting of metrics, and A/B testing and survey specifications.
Our experiments via oneline A/B testing and survey metrics showed a -1.4741\% to -2.273\% reduction in reports, and feedback survey submissions with user sentiment improvements of  -1.44\% to -3.9\% on key integrity areas.After we launched the survey model based our approaches on our platform, the model is able to bring reductions of 1.75\%, 2.57\%, 2.06\% on reports, dislikes, survey inappropriate rate, respectively. 

\section{Future Exploration}
In our current research, we acknowledge some limitations. For instance, when users receive surveys and provide their answers, they may respond randomly, which means that the results may not accurately reflect their true intentions. 
In future work, we will focus on addressing the issue of random responses in survey-based personalized modeling.

Furthermore, We will also explore the use of large recommendation models (LRM) that utilize negative feedback signals to enhance user experiences. 
To achieve this, our first step will be to diversify the types of negative feedback we collect. 
In addition to survey signals, we will include reports, dislikes, skips, and other forms of feedback. 
Secondly, we aim to optimize the LRM by employing more complex model structures and incorporating comprehensive sequential features. 
By leveraging the generalization capabilities of large models, we hope to effectively address the challenges presented by negative feedback signals and improve the overall user experience.


\bibliographystyle{ACM-Reference-Format}
\bibliography{sample-sigconf}

\end{document}